\documentclass[amsmath,amssymb,amsfont,aps,pra,twocolumn, superscriptaddress, nobibnotes,nofootinbib,longbib
]{revtex4-2}

\pdfoutput=1

\usepackage{enumitem}   
\usepackage{graphicx}
\usepackage{dcolumn}
\usepackage{bm}
\usepackage[colorlinks,citecolor=blue,linkcolor=blue,urlcolor=blue, breaklinks=true]{hyperref}
\usepackage{physics}
\usepackage{siunitx}  
\DeclareSIUnit\atomicmassunit{u}
\usepackage{xcolor}
\usepackage{ifthen}

\begin{document}

\title{
A truly relativistic gravity mediated entanglement protocol using superpositions of rotational energies
}

\author{Gerard Higgins}
\email{gerard.higgins@oeaw.ac.at}
\affiliation{Institute for Quantum Optics and Quantum Information (IQOQI), Austrian Academy of Sciences, A-1090 Vienna, Austria}
\affiliation{Department of Microtechnology and Nanoscience (MC2), Chalmers University of Technology, SE-412 96 Gothenburg, Sweden}
\author{Andrea Di Biagio}
\email{andrea.dibiagio@oeaw.ac.at}
\affiliation{Institute for Quantum Optics and Quantum Information (IQOQI), Austrian Academy of Sciences, A-1090 Vienna, Austria}
\affiliation{Basic Research Community for Physics e.V., Mariannenstraße 89, Leipzig, Germany}
\author{Marios Christodoulou}
\email{marios.christodoulou@oeaw.ac.at}
\affiliation{Institute for Quantum Optics and Quantum Information (IQOQI), Austrian Academy of Sciences, A-1090 Vienna, Austria}

\begin{abstract}
Experimental proposals for testing quantum gravity-induced entanglement of masses (QGEM) typically involve two interacting masses which are each in a spatial superposition state. Here, we propose instead a QGEM experiment with two particles which are each in a superposition of rotational states, this amounts to a superposition of mass through mass--energy equivalence. In sharp contrast to the typical protocols studied, our proposal is genuinely relativistic. It does not consider a quantum positional degree of freedom but relies on the fact that rotational energy gravitates: the effect we consider disappears in the $c\rightarrow\infty$ limit. Furthermore, this approach would test a feature unique to gravity since it amounts to sourcing a spacetime in superposition due to a superposition of `charge'.
\end{abstract}

\maketitle

Accessing empirically the regime in which quantum gravitational effects could take place is notoriously difficult. One way would, claimed impossible by Dyson \cite{dyson2012graviton,rothman2006can} would be to detect quantized radiation \cite{Tobar2023}.
Another avenue is to probe the gravitational field sourced by a quantum mass \cite{dewitt-morette2011role}. This idea has received much attention in recent years with the progress in quantum control over increasingly large masses \cite{aspelmeyer2014cavity,aspelmeyer2022when,weiss2021largea}, and well studied specific proposals for detecting QGEM that leverage quantum information concepts and trends in quantum technologies \cite{bose2017spin,marletto2017gravitationallyinduced,krisnanda2020observable,krisnanda2020observable}. Simple calculations using a direct Newtonian interaction show that two masses, each prepared in a position-delocalised state, will become entangled via the gravitational interaction. Linearised gravity predicts this effect \cite{christodoulou2023locally,marshman2020locality,bose2022mechanism}, while theories in which the gravitational interaction is mediated by a classical local field do not predict the generation of entanglement \cite{wallace2021quantum,oppenheim2023postquantum}. Thus, gravitationally mediated entanglement would be a signature of the non--classical nature of gravity. This is also supported by quantum information theoretic arguments that classical systems cannot mediate the creation of entanglement \cite{galley2022nogo,horodecki2009quantum,marletto2017witnessing}. While QGEM does not involve direct detection of gravitons, there is an interesting indirect connection between QGEM and quantised radiation within quantum field theory \cite{carney2022newton,mari2016experiments,belenchia2018quantum,danielson2022gravitationally}.

\begin{figure} 
\centering
\includegraphics[width=\columnwidth]{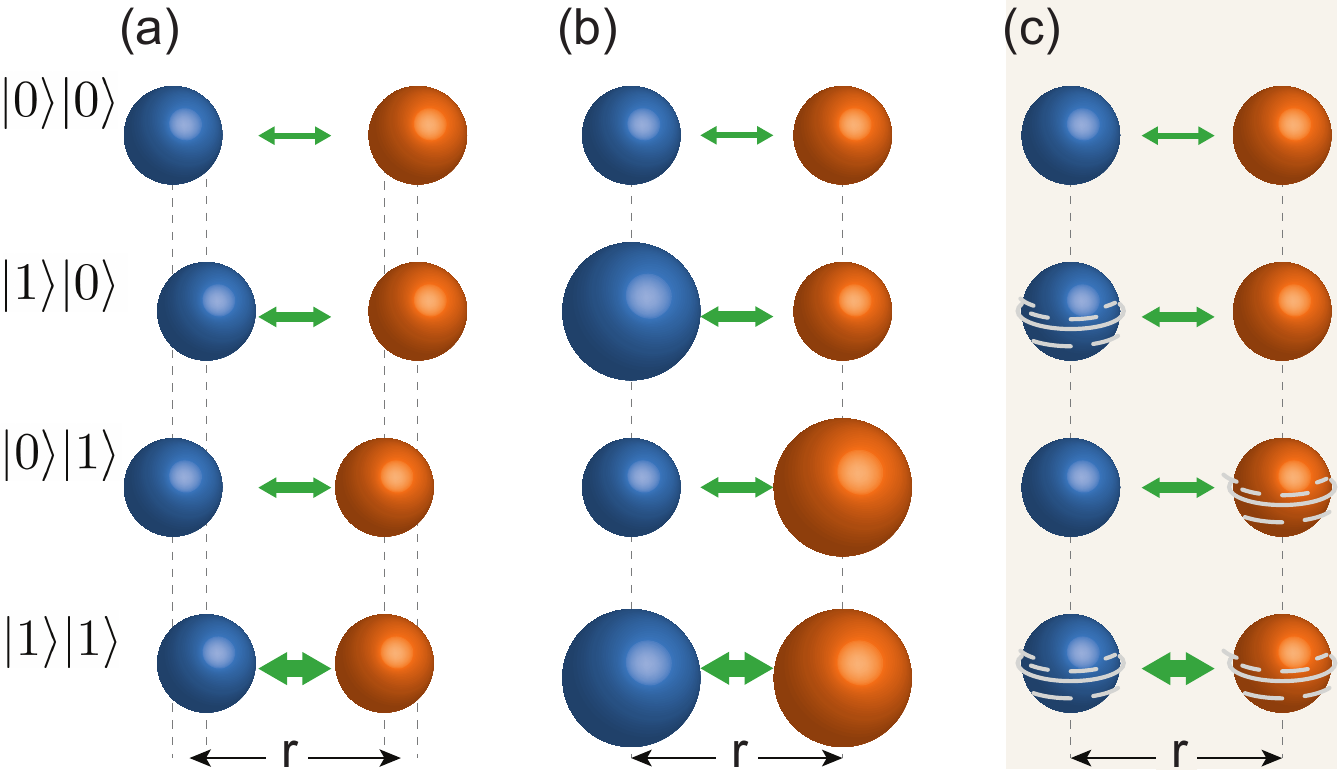}
\caption{
QGEM protocols aim to witness whether the gravitational interaction between two masses can be in a superposition state.
The typical approach considers (a)~preparing each mass in a superposition of locations. Another possibility is to (b)~prepare each particle in a superposition of mass-states. Concretely, we show howl this can be achieved by (c)~preparing each particle in a superposition of rotational energies and exploiting the equivalence between mass and (rotational) energy.
} \label{fig_two_balls}
\end{figure}

QGEM schemes are typically imagined in the Newtonian regime and concern sourcing the gravitational field with masses that are prepared so that their positional degree of freedom is in a quantum superposition. Here, instead, we study a relativistic QGEM scheme. Instead of considering spatial superposition states, we take two particles that are each in a superposition of states of \emph{different mass}, by preparing each particle in a superposition of rotational energy states and relying on mass-energy equivalence. In this case, not only $G$ and $\hbar$ are relevant but \emph{also} $c$: the effect disappears in the $c \rightarrow \infty$ limit. This idea is conceptually similar to the thought experiment discussed in Ref.~\cite{Castro_Ruiz_2017}, where it is shown that two clocks will get entangled due to the fact that a clock generally involves transitions between different energy states, and in Ref.~\cite{marletto2018quantumgravity}, where neutrino-like oscillations are considered.
QGEM protocols using particles in spatial superposition states and in superpositions of mass states are schematically represented in Fig.~\ref{fig_two_balls}.

Performing such an experiment would be a more important test than the usual QGEM protocols that operate in the Newtonian regime. The protocol we study here relies on the general relativistic effect whereby the gravitational field can be sourced by rotational energy. This effect does not take place in Newtonian gravity, where the source of the gravitational field is the mass density. Because the use of the mass--energy results in a $1/c^4$ suppressing factor in the entangling phase, this test would be a more difficult experiment than detecting QGEM with spatial superposition states. 
However, this protocol would test a genuine interface between general relativity and quantum mechanics in the sense that there is no precise analogue in electromagnetism as it would amount to considering a particle in a state superposition of electric charge which is excluded by superselection rules. We discuss the importance of this protocol in testing whether gravity can be treated as a classical variable in our concluding section. 

Coming up with a realistic protocol to detect QGEM in mass superposition states is therefore not a trivial matter. This is the task for the current work. The protocol we discuss below goes some way in bringing this possibility to the realm of the feasible. Indeed, while we identify an operational `table top' parameter regime, a sober headed conclusion would be that the parameters we use are extremely ambitious and are not feasible for the foreseeable future.

We start with some general considerations of the difficulties in performing such an experiment. Then we consider a specific protocol for achieving a macroscopic superposition of rotational energies. We close with a discussion of our findings. 

\section{General considerations}
The Hamiltonian for two particles interacting via gravity at a fixed separation $r$ is $\hat{H} = -{G \hat{M}_1 \hat{M}_2}/{r}$ where $G$ is Newton's constant, $\hat{M}_i$ is the mass operator acting on particle $i$ \cite{Zych2017}.
We take each particle being initialised in an equal superposition of mass $M$ and mass $M+\Delta M$, denoted as $\ket0$ and $\ket1$, ${\ket{\Psi(0)} = \frac{1}{2} \left( \ket{0} + \ket{1} \right) \left( \ket{0} + \ket{1} \right)}$. 
After time $T$ the state of the system can be written as ${\ket{\Psi(T)} = \frac{1}{2} \left( \ket{00} + \ket{01} + \ket{10} + e^{i \phi} \ket{11} \right)}$, where ${\phi =  \frac{G}{\hbar} \frac{\Delta M^2 T}{r}}$ (the calculation is described step by step in \cite{Supplemental}).
This two-qubit state is entangled if $\phi \neq 0$. The evolution of the two masses interacting via gravity can be interpreted as an entangling two-qubit controlled-phase gate, with $\phi$ the controlled phase shift \cite{Polino:2022qxg}.
The degree of entanglement of the state $|\Psi(T)\rangle$ can be quantified by the concurrence $|\sin{2\phi}\,|$.

If the mass superposition $\Delta M$ is achieved by preparing each particle in a superposition of energy states, with energy difference $E$, the entangling phase is
\begin{equation} \label{eq_phi_E}
    \phi = \frac{G}{\hbar c^4} \frac{E^2 T}{r}
\end{equation}
Note the suppressing factor $1/c^4$ in the entangling phase which is due to the use of the mass--energy equivalence when considering a gravitating energy as we do here. If $\phi$ is small, then with $N$ experimental repetitions, the statistical uncertainty in an estimate of $\phi$ is $\sim 1/\sqrt{N}$.
To show that $\phi$ is non-zero, and that there is entanglement in the system, $N \sim 1/\phi^2$ experiment repetitions are needed.

A back-of-the-envelope estimate gives an initial idea of the challenging requirements: With evolution time $T=\SI{1}{\second}$, $N=10^8$ repetitions (requiring a minimum total experiment time of $\sim$ 3 years), and separation $r=\SI{1}{\micro\meter}$, entanglement would become measurable if $\Delta M = \SI{4E-18}{\kilogram} = \SI{2E9}{\atomicmassunit}$ and ${E = \SI{0.4}{\joule}}$: a truly macroscopic energy superposition would be needed! 

For example, it is unrealistic to consider the use of individual atoms or molecules in superpositions of electronic states, as these are separated by ${E \sim \SI{1}{\eV} \approx \SI{E-19}{\joule}}$. Similarly, if we assume that a large number of atoms or molecules are probed in parallel and long coherence times (record coherence times exceed 1 hour \cite{Wang2021}), the phase remains unresolvable. The same applies for superpositions of nuclear states, since the fusion and fission processes typically involve energies ${\sim \SI{e8}{\eV} \sim \SI{E-11}{\joule}}$.  

A more realistic approach to creating superposition of energy--mass with a view to testing QGEM is to consider two macroscopic solid massive rotors, each of which is set in a superposition of rotational energy states---and thus a superposition of masses---interacting via gravity. 

A rotor with moment of inertia $I$ in a superposition of rotating with angular velocity $\omega$ and angular velocity $0$ is in a superposition of rotational energies with energy difference $E=\tfrac{1}{2}I\omega^2$.
Substituting this energy difference into Eq.~(\ref{eq_phi_E}), the entangling phase which develops is $\phi = {G I^2 \omega^4 T}/({4 \hbar c^4 r})$.
The general idea studied here can be thought of as investigating whether the quartic dependence of $\phi$ on the angular velocity $\omega$ can be used to offset the suppressing factor $1/c^4$. As we will see, there are several other trade--offs that need to be considered in a realistic analysis.

For example, optically-levitated nanoparticles \cite{Kuhn2017} have indeed been spun-up to very high GHz rotational frequencies using circularly-polarized light \cite{Ahn2018, Reimann2018}. However,
this approach is not suitable due to the unwanted absorption of photons. 

Instead, we consider two rotors, each with an embedded electric dipole moment and an embedded magnetic dipole moment, which can be controlled using electric fields and magnetic fields.
We remark that using rotational degrees of freedom for gravitational entanglement has been proposed in \cite{pedernales2020decoherencefree}, where a superposition of different static orientations of complex shapes leads to entanglement, while the more recent \cite{lantao2024lowenergy} proposes to entangle angular momentum degrees of freedom via the Lens-Thirring effect.
For an introduction to the topic of spin-controlled rotors, see for instance the proposals in Refs.~\cite{Rusconi2022, Ma2021} and the experiments described in Refs.~\cite{Urban2019, Delord2020, Jin2023}. For a review of quantum phenomena that appear with rotating nanoparticles see Ref.~\cite{Stickler2021}.

\section{Protocol}
We consider two solid particles, each has a large magnetic dipole moment, as well as an embedded spin-1/2 particle corresponding to an electric dipole moment
The idea is to use electric control of the relatively weak spin-1/2 electric dipole moment to prepare a superposition of orientations, then to use magnetic control of the relatively large magnetic dipole moment to spin-up each particle.
We describe the overall state of each particle in terms of the orientation of the spin-1/2 electric dipole moment $\{|0\rangle, |1\rangle \}$ and the orientation of the rest of the particle (which includes the embedded magnetic dipole moment) in terms of an angle $\theta$ and angular velocity $\omega$, $|\theta, \omega\rangle$.
We consider initiating each particle in state $|0\rangle \otimes |\theta=0, \omega=0\rangle$ and then applying the following steps:

\begin{enumerate}[label=(\arabic*)]
    \item Prepare each spin-1/2 electric dipole moment in a superposition of orientations. After this step, each particle is in state $\frac{1}{\sqrt{2}} \left( |0\rangle + |1\rangle 
 \right) \otimes |\theta=0, \omega=0\rangle$.
    \item Apply an electric field, to cause each particle to rotate to a superposition of orientations. This step lasts for time $T_2$. This results in the large embedded magnetic dipole moment having a superposition of orientations. After this step, each particle is in state $\frac{1}{\sqrt{2}} \left( |0\rangle \otimes |\theta=-\theta_0, \omega\approx0\rangle + |1\rangle 
  \otimes |\theta=\theta_0, \omega\approx0\rangle \right)$.
    \item Apply an alternating magnetic field, to spin up each particle. The field is orientated along $-\theta_0$, so it effects a torque on only the $|1\rangle$ component of the superposition, and not on the $|0\rangle$ component. This step lasts for time $T_3$. After this step, each particle is in state $\frac{1}{\sqrt{2}} \left( |0\rangle \otimes |\theta=-\theta_0, \omega\approx0\rangle  + |1\rangle \otimes |\theta(t), \omega=\omega_\mathrm{max}\rangle 
   \right)$
    \item Allow the two particles to interact via gravity for time $T_4$.
    \item Reverse step (3).
    \item Reverse step (2).
    \item Measure.
\end{enumerate}
Let us describe some of the steps in more detail:

{Step (2):}
If the electric dipole moment states are initiated along the $\pm \hat{x}$ directions, application of an electric field along the $\hat{y}$ direction will cause a torque about the $\pm \hat{z}$ direction.
This torque will have magnitude $p \mathcal{E}$, where $\mathcal{E}$ is the magnitude of the electric field.
By the end of step~(2) each component will be reorientated by an angle $\theta_0 \approx {p \mathcal{E} T_2^2}/{(2I})$,
where $T_2$ is the duration of this step and $I$ is the particle's moment of inertia.

\begin{figure} 
\centering
\includegraphics[width=0.75\columnwidth]{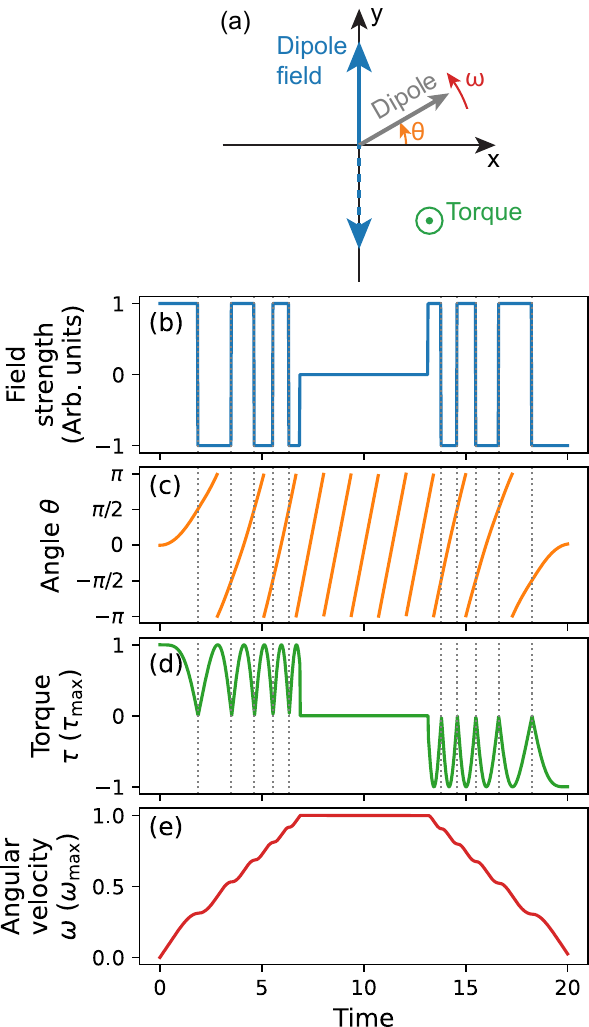}
\caption{A dipole field can be used to spin-up a dipole moment.
The dipole moment in the $x-y$ plane has polar angle $\theta$.
When $\pi/2<\theta<3\pi/2$ a field along $-\hat{y}$ is applied, otherwise a field along $\hat{y}$ is applied.
This ensures the torque is always about $\hat{z}$.
The process can be reversed.
} \label{fig_spin_up}
\end{figure}

{Step (3):}
A magnetic field then applies a torque on the $|1\rangle$ component and not on the $|0\rangle$ component.
This is achieved by applying the field paralllel with the orientation of the magnetic dipole moment of the $|0\rangle$ component.
Fig.~\ref{fig_spin_up} illustrates how the field can be alternated as the particle rotates to ensure that the torque is always about $\hat{z}$, and that only the $|1\rangle$ component is spun up.

With a magnetic field with strength $B$ acting on the magnetic dipole moment $m$ the maximum torque is $\tau_\mathrm{max}=mB$ (note that $m$ refers to a magnetic moment not a mass).
The average torque during step~(3) is approximately $\tfrac{2}{\pi}\tau_\mathrm{max}$ [c.f.\ Fig.~\ref{fig_spin_up}(e)].
The angular momentum of the $|1\rangle$ component will evolve to ${I \omega_\mathrm{max} = \tfrac{2}{\pi}\tau_\mathrm{max} T_3 = \tfrac{2}{\pi} mB T_3}$.
To a good approximation, the angular velocity of the $|1\rangle$ component will increase linearly up to $\omega_\mathrm{max}$ during step (3), remain constant during step (4), then decrease linearly during step (5), as shown in Fig.~\ref{fig_spin_up}(d).

The evolution of the angular velocity during step~(3) depends on the separation between the orientations $2\theta_0$ at the end of step~(2).
The evolution will proceed fastest if $2\theta_0=\pi/2$, this case is represented in Fig.~\ref{fig_spin_up}.
The more general cause, for lower values of $\theta_0$, is described in \cite{Supplemental}.

\begin{figure} [ht!]
\centering
\includegraphics[width=0.9\columnwidth]{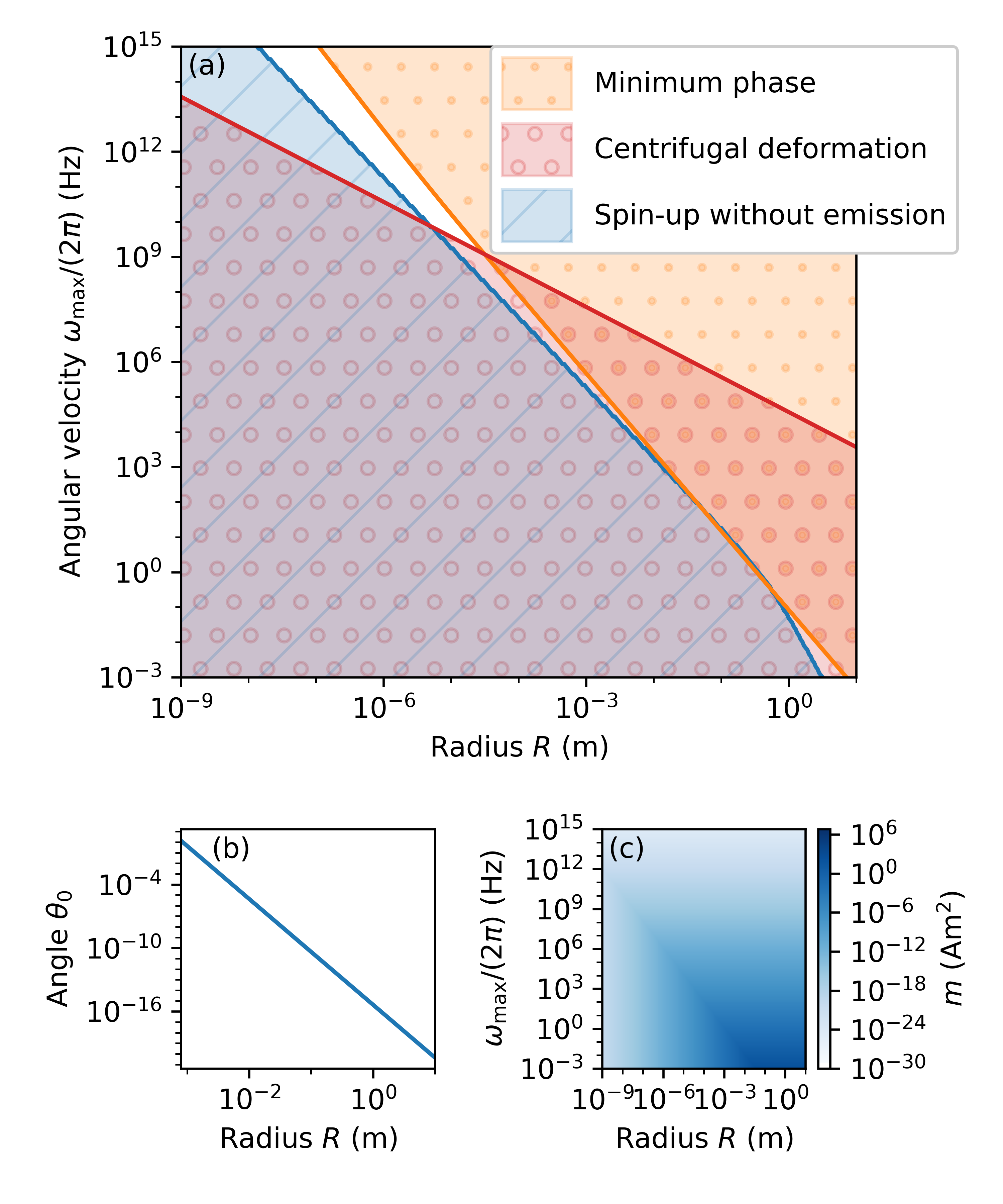}
\caption{(a)~The shaded regions represent parameter regimes which satisfy the different requirements we consider. The shaded regions overlap for particles with radii around $\SI{0.1}{\meter}$ radius and angular velocities around $2\pi\times\SI{1}{\hertz}$, indicating the scheme may be viable for these parameters.
(b)~It becomes more difficult to prepare superpositions of orientations (separated by angle $\theta_0$) during step~(2) as the particle size and the moment of inertia is increased.
(c)~The magnetic dipole moment $m$ is limited by the particle volume and remanence magnetisation for small particles, and by the requirement that photons are not emitted for fast-spinning particles.
These figures consider using two solid spheres.
} \label{fig_scheme2}
\end{figure}

In the next section, we will consider different effects which place limits on the protocol.
Parameter regimes allowed by these effects are indicated in Fig.~\ref{fig_scheme2}(a).
We consider using two solid spheres.
The main parameters are the sphere radii $R$, maximum angular velocity $\omega_\mathrm{max}$, and the duration of each step (we optimistically consider $10^3\,\mathrm{s}$).
We also consider particles of density $\SI{2.3E4}{\kilogram\per\meter\cubed}$, which is the highest atmospheric pressure density of any element (osmium). 

\section{Operational limitations}
We now consider three limitations and how to overcome them: the smallness of the entangling phase, centrifugal deformation, and spinning up without decoherence. 

\subsection{Entangling phase}
The entangling phase $\phi$ that will develop over steps (2)-(4) is $\phi = \frac{G}{4 \hbar c^4} \frac{I^2}{r} \int  \omega(t)^4  dt \approx \frac{G}{4 \hbar c^4} \frac{I^2 \omega_\mathrm{max}^4}{r} \left( \frac{2 T_3}{5} + T_4 \right)$.
We aim for phase ${\phi > 10^{-3}}$, to avoid requiring too many experimental repetitions for the phase to be resolved.
Furthermore, very small values of $\phi$ would be difficult to measure reliably due to systematic uncertainties.
This criterion is satisfied in the upper right region of Fig.~\ref{fig_scheme2}(a) (orange dotted area), for higher values of $\omega_\mathrm{max}$ and higher particle radii $R$.

\subsection{Centrifugal deformation}
Fast spinning objects can irreversibly deform due to centrifugal forces.
This happens when the tangential speed becomes comparable with the speed of sound in the material, introducing distinguishability between the spinning and non-spinning branches (components $|1\rangle$ and $|0\rangle$ respectively), thus causing decoherence.
This is avoided in the lower left region of Fig.~\ref{fig_scheme2}(a) (red region with open circles), for lower values of $\omega_\mathrm{max}$ and lower particle radii $R$.
For the speed of sound in the particles, we use the highest physically-allowed value of $\SI{3.7E4}{\meter\per\second}$ \cite{Trachenko2020}.

\subsection{Spin-up without radiation emission}
The third limitation we consider is somewhat more involved than the two introduced so far.
It deals with whether it is possible to spin-up the particle to $\omega_\mathrm{max}$ while avoiding decoherence caused by radiation.

With this in mind, first we need to consider the size of the superposition of orientations $\theta_0$ achieved after step (2).
The electric field strength we can consider applying during step (2) is $\mathcal{E} \approx 10^{10}\,\mathrm{V m^{-1}}$, since matter is ionised in higher electric field strengths.
We consider having control over electric dipole moments $p=10^3\,\mathrm{D}$, dipole moments of this magnitude were prepared in superposition states using laser light and microwave radiation in Ref.~\cite{Zhang2020}.
This gives a maximum torque $p\mathcal{E} \approx 10^{-17}\,\mathrm{Nm}$ during step (2).
Fig.~\ref{fig_scheme2}(b) shows how the angle $\theta_0$ that can be achieved decreases as the particle's moment of inertia is increased.

Then, whether or not the particle of radius $R$ can be spun-up to $\omega_\mathrm{max}$ depends on the magnitude of the embedded magnetic moment $m$, the strength of the applied magnetic field $B$, as well as on the size of the angular superposition $\theta_0$ prepared during step (2).

Limits on the magnetic moment $m$ are represented in Fig.~\ref{fig_scheme2}(b).
If the particles are small, $m$ will be restricted by the particle volume $V$ and the remanence magnetisation $M_\mathrm{r}$ to $m=M_\mathrm{r}V$.
We consider the high magnetisation of neodymium $M_\mathrm{r}=\SI{1.4}{\tesla}$.
If the particles spin with high angular velocities, $m$ will be limited by the need to avoid emission of radiation, which we explain next.

The rotating magnetic dipole moment may emit electromagnetic radiation.
If the rotating component of the superposition (state $|1\rangle$) emits a photon, then states $|0\rangle$ and $|1\rangle$ will be distinguishable, and the coherence between them will be lost.
And so, the scheme requires there to be a low probability of photon emission.

The probability of emitting one or more photons is given bythe overlap between the vacuum and the coherent state of the electromagnetic field sourced by the spinning branch, which is $e^{-\langle n\rangle}$, where $\langle n\rangle$ is the expectation number of the emitted photons. A classical magnetic dipole moment rotating with angular velocity $\omega$ radiates photons of energy $\hbar \omega$. The power emitted is given by $P = {\omega^4 m^2}/{(6 \pi \epsilon_0 c^5})$, the rate of photon emission is $\dot{n} = {\omega^3 m^2}/({6 \pi \epsilon_0 \hbar c^5})$ and so, during steps (3-5), the expectation number of emitted photons is $\langle n\rangle = {\omega_\mathrm{max}^3 m^2}/({6 \pi \epsilon_0 \hbar c^5}) \left(\frac{T_3}{3} + T_4 \right)$.
We limit $m$ by requiring $\langle n \rangle < 1$.
This limit impacts $m$ in Fig.~\ref{fig_scheme2}(c) if $\omega_\mathrm{max}$ is relatively high.


Then, the parameter regime which allows the particles with radii $R$ and magnetic moment $m$ to be spun up to $\omega_\mathrm{max}$ is indicated by the parameter regime with the blue striped lines in Fig.~\ref{fig_scheme2}(a).
This region is in the lower left part of the figure, since it is easier to spin-up smaller particles with smaller moments of inertia to smaller angular velocities.
Here we considered using a magnetic field with strength $B=\SI{100}{\tesla}$ (this is around twice the record field strength that has been achieved in continuous operation \cite{Miller2003,Hahn2019}).

\subsection{Overcoming the limitations}
It appears to be possible to satisfy these three requirements for particles around $R\sim\SI{0.1}{\meter}$ and angular velocities $\omega_\mathrm{max} \sim 2\pi \times \SI{1}{\hertz}$, since the three shaded regions in Fig.~\ref{fig_scheme2}(a) overlap around these parameters.

The overlap of the shaded regions improves if longer durations are allowed, however, the timescales we consider ($10^3\,\mathrm{s}$) are already extremely ambitious, since the experiment will be affected by other sources of decoherence \cite{Zhong2016, Stickler2018}.
These include blackbody radiation \cite{Chang2010}, collisions with background gas particles \cite{Chang2010, Martinetz2018} and surface interactions between the rotors \cite{Martinetz2022}.
Achieving quantum coherence times around $10^3\,\mathrm{s}$ for solid particles with radii around $0.1\,\mathrm{m}$ is far, far beyond the current state of the art.
Low temperatures and low background gas pressures many orders of magnitude beyond the current state of the art would be needed to mitigate these decoherence mechanisms.

As well as the parameters described so far, we considered a separation $r=2R+r_\mathrm{min}$ between the spheres, where the minimum separation $r_\mathrm{min}=\SI{10}{\micro\meter}$.
While lower separations allow for stronger gravitational interactions, electric and magnetic interactions between the particles will need to be shielded. The value of $r_\mathrm{min}$ is chosen to allow for the possibility of an electromagnetic shield between the rotors.

Additional effects that arise due to trapping the rotors \cite{Martinetz2021_NJP} are outside the scope of this work.
A space-based experiment \cite{kaltenbaek2022maqro} could avoid the need for trapping, but this would raise other technical challenges.

One can consider different avenues for relaxing the experimental requirements.
By reducing the separation between the masses, stronger gravitational interactions can be achieved. And so using two disc-shaped particles should enable larger operational parameter regimes than using two spheres, as we show in \cite{Supplemental}.
 
\section{Conclusion}

We have studied the possibility of using superposition of masses to detect entanglement mediation through gravity. In contrast to the typical QGEM proposals, this protocol is not described by Newtonian gravity. The superposition of mass is achieved through the use of superposition of rotational states and making use of the equivalence of rotational energy with mass. This setup for detecting mediated entanglement is also unique to gravity as the electromagnetic analogue would involve a superposition of charge. Entanglement arises because rotational energy gravitates in general relativity, an effect that does not take place in Newtonian gravity (the effect dissappears if we take $c \rightarrow \infty$). 

The importance of detecting gravity mediated entanglement has been extensively discussed in the literature. The seminal papers \cite{bose2017spin,marletto2017gravitationallyinduced} (see also \cite{marletto2017witnessing, galley2022nogo}) presented no-go theorems stating that detection of gravity mediated entanglement that implies that, in any \textit{local} theory, gravity must be mediated by a non-classical system. These arguments are generalisations to post-quantum theories of the well known fact that local operations and classical communications cannot increase entanglement \cite{horodecki2009quantum}.
However, it is crucial to note that these theorems assume locality at the level of \textit{subsystems}, which is a distinct notion from spacetime locality. While the latter is a well established notion of locality in field theory, the subsystem notion of locality is on much weaker grounds, as one can have relativistically local theories with or without subsystem-local interactions \cite{DiBiagio:2023jiu,fragkos2023inference,Christodoulou:2022knr}.

A much less ambiguous way to state why these experiments are important is via the standard hypothesis testing route \cite{huggett2023quantum}: some theories predict the effect and some do not and performing the experiment will allow us to distinguish between them.

Within linearised quantum gravity, the detection of mediated entanglement may be said to evidence that the gravitational field can be set in a quantum superposition of diffeomorphically inequivalent configurations~\cite{Christodoulou:2018cmk, christodoulou2023locally,Martin-Martinez:2022uio}.

Our protocol admits a similar interpretation as the usual GME protocol, with the main difference of going one step beyond the Newtonian limit of linearised quantum gravity, by witnessing the genuinely relativistic effect of energy sourcing the gravitational field. The relativistic nature of this effect is manifested in the strong suppressing factor $1/c^4$ in the entangling phase [Eq.~(\ref{eq_phi_E})].

Another interesting distinction between  our proposal and the spatial superposition QGEM experiments is that the effect has no electromagnetic analog, as it is well-known that is not possible to create a superposition of electric charges.

Our scheme involves achieving superpositions of rotations using control over electric dipole moments and magnetic dipole moments.
The scheme gives a small parameter regime which may be operational, see Fig.~\ref{fig_scheme2}.
In order to achieve a workable regime, we had to make quite ambitious choices of parameters. While further improvements to our scheme are conceivable  that may relax these parameters, we expect them to remain ambitious with respect to the current state of the art. We conclude that while it is conceivable to realise the task at hand in a `table top' setup, it would present a formidable experimental challenge and require significant technological improvements.

\section*{Acknowledgments}
We thank Benjamin Stickler, Corentin Gut and Markus Aspelmeyer for fruitful discussions.
G.H.\ acknowledges support from the Swedish Research Council (Grant No.\ 2020-00381).
M.C. and A.D.B. acknowledge support of the ID~\#~61466 grant from the John Templeton Foundation, as part of the ``Quantum Information Structure of Spacetime (QISS)'' project (\hyperlink{http://www.qiss.fr}{qiss.fr}).
This research was funded in whole or in part by the Austrian Science Fund (FWF) [10.55776/esp525]. For open access purposes, the authors have applied a CC BY public copyright license to any author-accepted manuscript version arising from this submission.


%

\renewcommand{\theequation}{S\arabic{equation}}
\renewcommand{\thefigure}{S\arabic{figure}}

\section*{Supplemental material}

\section{Entangling phase calculation} \label{appendix_entangling_phase_calculation}
The Hamiltonian describing two particles interacting via gravity is
\begin{equation}
    \hat{H} = -\frac{G \hat{M}_1 \hat{M}_2}{r}.
\end{equation}
where $G$ is Newton's constant, $\hat{M}_i$ is the mass operator acting on particle $i$, and $r$ is the separation. We take each particle being initialised in an equal superposition of mass $M$ and mass $M+\Delta M$, denoted as $\ket0$ and $\ket1$ 
\begin{align}
\begin{split}
    \ket{\Psi(0)} &= \frac{1}{2} \left( \ket{0} + \ket{1} \right) \left( \ket{0} + \ket{1} \right).
\end{split}
\end{align}
After time $T$ the system evolves to the state
\begin{align}
    \ket{\Psi(T)} & = \frac{1}{2} ( e^{i \phi_{00}} \ket{00} + e^{i \phi_{01}} \ket{01}  \nonumber \\ &+ e^{i \phi_{10}} \ket{10} + e^{i \phi_{11}} \ket{11} )
\end{align}
where
\begin{align}
\begin{split}
    \phi_{00} &= \frac{G M^2 T}{\hbar r} \\
    \phi_{01} = \phi_{10} &= \frac{G M(M+\Delta M) T}{\hbar r} \\
    \phi_{11} &= \frac{G (M+\Delta M)^2 T}{\hbar r}.
\end{split}
\end{align}
and if any evolution of the particle positions is negligible. Changing basis, the state of the system at time $T$ can be written as
\begin{equation}
    \ket{\Psi(T)} = \frac{1}{2} \left( \ket{00} + \ket{01} + \ket{10} + e^{i \phi} \ket{11} \right)
\end{equation}
with the phase
\begin{align}
\begin{split}
    \phi &= \phi_{00} + \phi_{11} - \phi_{01} - \phi_{10}=  \frac{G}{\hbar} \frac{\Delta M^2 T}{r}
\end{split}
\end{align}
This two-qubit state is entangled if $\phi \neq 0$. The evolution of the two masses interacting via gravity can be interpreted as an entangling two-qubit controlled-phase gate, with $\phi$ the controlled phase shift. 
The degree of entanglement of the state $|\Psi(T)\rangle$ can be quantified by the concurrence $|\sin{2\phi}\,|$.

\section{Dynamics during step (3)} \label{appendix_spin_up}
\begin{figure} [ht]
\centering
\includegraphics[width=\columnwidth]{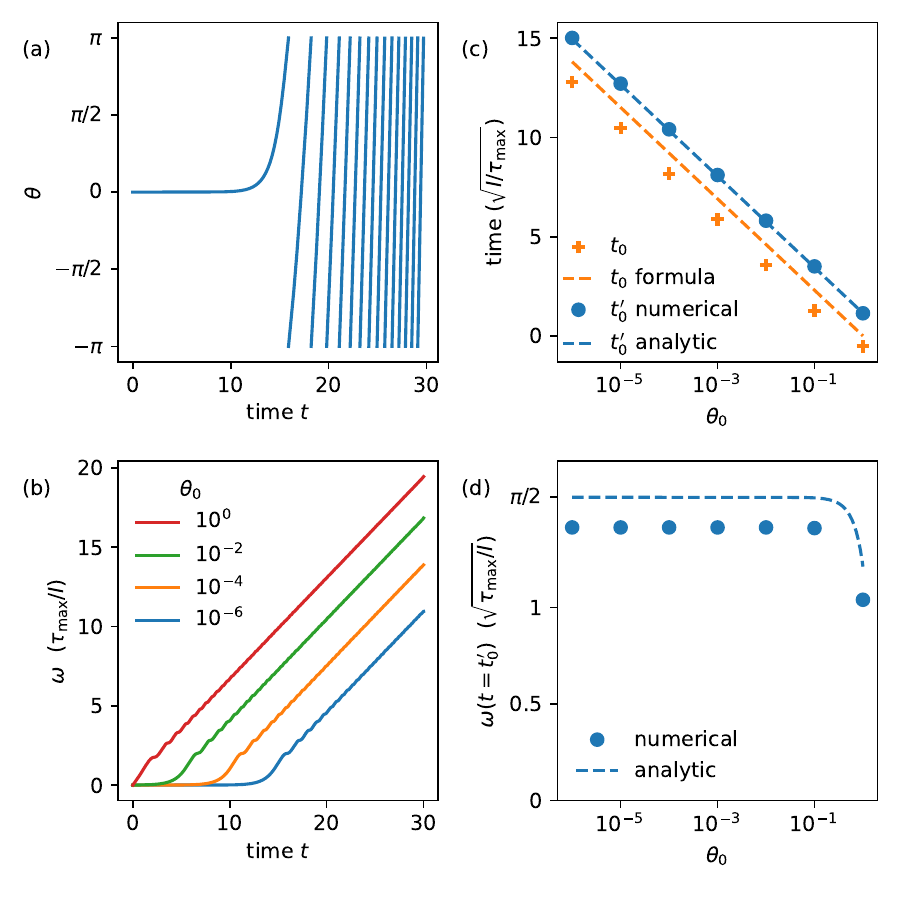}
\caption{Dynamics during ramp-up (step 3) are slower than described in the main text.
(a)~Ramp-up (showing $\theta$) for initial angle $\theta_0=10^{-6}$.
(b)~Ramp-up (showing $\omega=\dot{\theta}$) for different initial angles. The curves are flattish until time $t_0$. Smaller $\theta_0$ values give longer $t_0$ values.
(c)~The inverse relation between $t_0$ and $\theta_0$ is shown more clearly. The time $t_0'$ at which $\theta$ reaches $\pi/2$ behaves similarly to $t_0$.
(d)~The angular velocity at time $t_0'$ is insensitive to $\theta_0$.
} \label{fig_slowed_ramp_up}
\end{figure}

If, at the end of step 2, the angular separation $\theta_0$ between the magnetic dipole moments in the two branches ($|0\rangle$ and $|1\rangle$) is $\pi/2$, then during step 3 the angular velocity will grow (approximately) linearly in time, just as in Fig.~2(e).
On the other hand, if $\theta_0 < \pi/2$, the evolution during step 3 will be slowed, since the torque that is applied on the $|1\rangle$ branch will be initially only $\tau_\mathrm{max} \sin{\theta_0}$.
This is illustrated in Fig.~\ref{fig_slowed_ramp_up}(a), for $\theta_0=10^{-6}$.

We investigate the dynamics using numerical integration, and find the evolution during step 3 can be well-described by
\begin{equation} \label{eq_omega_empirical}
    \omega = \begin{cases} 0 & t<t_0 \\
    \frac{2}{\pi} \frac{\tau_\mathrm{max}}{I} (t-t_0) & t \geq t_0
    \end{cases}
\end{equation}
where
\begin{equation} \label{eq_t0}
    t_0 = \sqrt{\frac{I}{\tau_\mathrm{max}}} \ln{\frac{1}{\theta_0}}
\end{equation}
This is shown in Fig.~\ref{fig_slowed_ramp_up}.
The blue line representing the dynamics of the angular velocity $\omega$ in Fig.~\ref{fig_slowed_ramp_up}(b) is the gradient of the curve showing the dynamics of the angle $\theta$ in Fig.~\ref{fig_slowed_ramp_up}(a).
The different lines in Fig.~\ref{fig_slowed_ramp_up}(b) consider different starting $\theta_0$ values.
The $t_0$ value is extracted by fitting each of these curves by Eq.~(\ref{eq_omega_empirical}), then the dependence of $t_0$ on $\theta_0$ is plotted in Fig.~\ref{fig_slowed_ramp_up}(c).
As shown in the figure, this dependence is well described by Eq.~(\ref{eq_t0}).

We used Eqs.~(\ref{eq_omega_empirical}) and (\ref{eq_t0}) to produce Figs.~3 and \ref{fig_scheme2_discs}.

In the rest of this appendix we use analytics to understand how the dynamics described by Eqs.~(\ref{eq_omega_empirical}) and (\ref{eq_t0}) arise.
Readers who are content with an empirical formula based on numerical calculations do not need to read this appendix further.

The torque depends on the rotor orientation
\begin{align} \label{eq_differential2}
    \tau = I \ddot{\theta} &= \tau_\mathrm{max} |\sin{\theta}| \\
    &\approx \tau_\mathrm{max} \theta \label{eq_differential3}
\end{align}
where the approximation is valid for small angles.

The differential equation in Eq.~(\ref{eq_differential2}) does not have an analytical solution, whereas the differential equation in Eq.~(\ref{eq_differential3}) has an analytical solution.
For much of the dynamics $\theta$ is small, which allows us to focus on the solution of Eq.~(\ref{eq_differential3}), which is 
\begin{equation} \label{eq_theta_cosh}
    \theta(t) = \theta_0 \cosh{\sqrt{\frac{\tau_\mathrm{max}}{I}}t}
\end{equation}
for $\theta(t=0)=\theta_0$ and $\omega(t=0)=0$.

We define the time $t_0'$ as the time it takes for the rotor to rotate by $\pi/2$
\begin{align}
    t_0' &= \sqrt{\frac{I}{\tau_\mathrm{max}}} \mathrm{arccosh}{\frac{\pi}{2 \theta_0}} \label{eq_t0'_analytical1} \\
    &\approx \sqrt{\frac{I}{\tau_\mathrm{max}}} \ln{\frac{\pi}{\theta_0}} \label{eq_t0'_analytical2}
\end{align}
where the approximation is valid for $\theta_0 \ll \pi$.
This behaviour is similar to that of Eq.~(\ref{eq_t0}).
$t_0'$ is a good proxy for $t_0$, as shown in Fig.~\ref{fig_slowed_ramp_up}(c).
This is unsurprising, since when $\theta=\pi/2$, the rotor experiences the maximal torque and it accelerates relatively quickly, and then $\omega$ should be described by the linearly-increasing regime of Fig.~\ref{fig_slowed_ramp_up}(b).
Fig.~\ref{fig_slowed_ramp_up}(c) displays both the analytical values of $t_0'$ [from Eq.~(\ref{eq_t0'_analytical2})] and the values of $t_0'$ found by numerically solving Eq.~(\ref{eq_differential2}).

Lastly, as a final check, in Fig.~\ref{fig_slowed_ramp_up}(d) we see $\omega(t=T)$ is largely independent of $\theta_0$.
This follows from Eqs.~(\ref{eq_theta_cosh}) and (\ref{eq_t0'_analytical1})
\begin{align}
    \omega(t=t_0') &= \sqrt{\frac{\tau_\mathrm{max}}{I}} \theta_0 \sinh{ \mathrm{arccosh} \frac{\pi}{2\theta_0} } \\
    &\approx \sqrt{\frac{\tau_\mathrm{max}}{I}} \frac{\pi}{2}
\end{align}
where the approximation is valid for small $\theta_0$.

\section{Disc-shaped particles} \label{appendix_discs}
\begin{figure} [ht]
\centering
\includegraphics[width=\columnwidth]{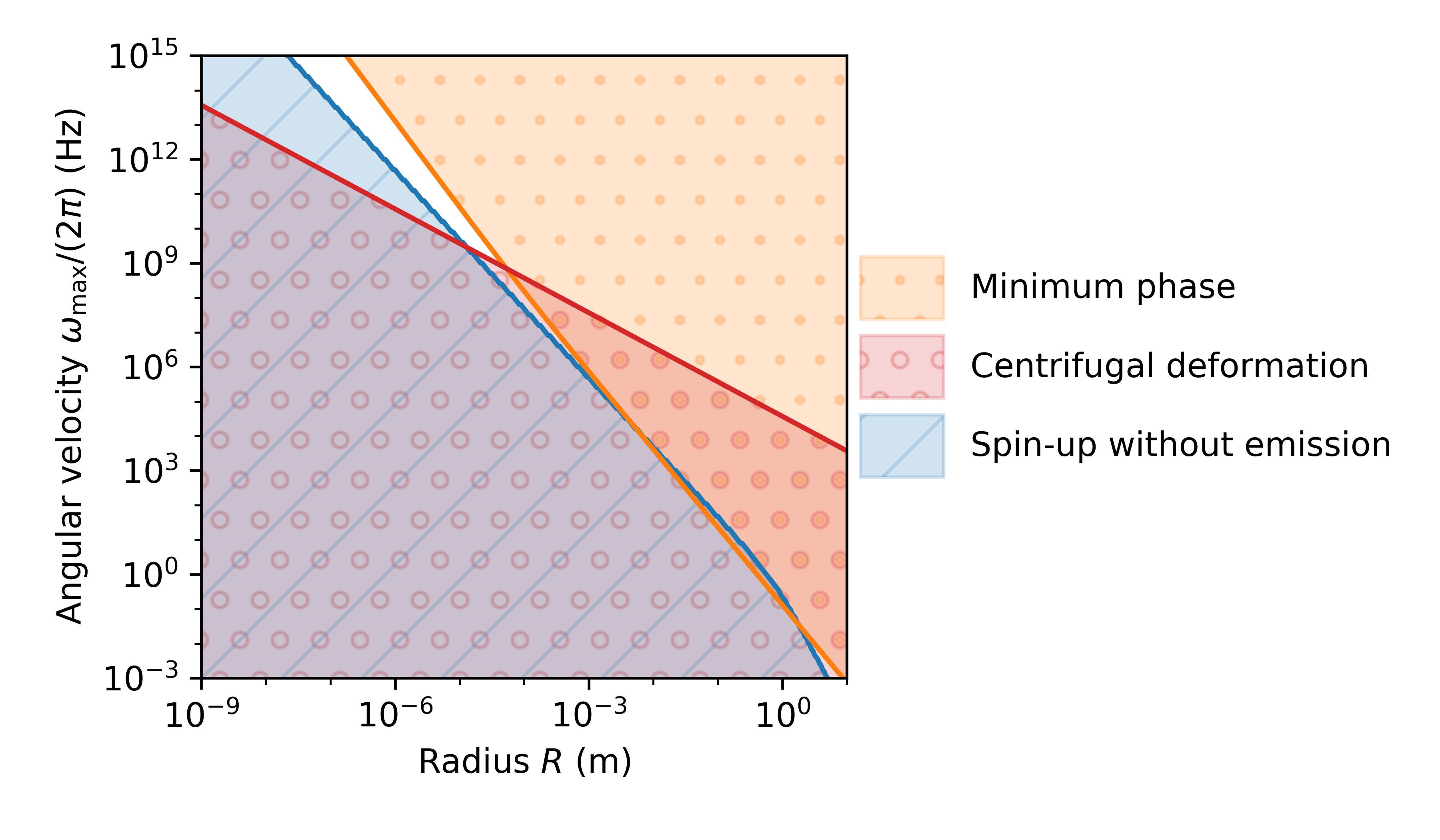}
\caption{Conducting the protocol using disc-shaped particles enlarges the overlap between the regimes satisfying the requirements we studied.
} \label{fig_scheme2_discs}
\end{figure}

The simplest case is to consider disc-shaped particles rather than spheres.
We repeat the above analysis for two discs orientated face-to-face, each with radius $R$ and height $H=R/10$, separated by $r=H+r_\mathrm{min}$.
The other parameters are kept the same as for the two spheres considered above.
Fig.~\ref{fig_scheme2_discs} shows the results.
Note that we now have a larger operational parameter regime compared to Fig.~3(c).
The situation will improve further if the ratio $H/R$ is further reduced. 

\end{document}